\documentclass[prx,twocolumn,superscriptaddress,nofootinbib]{revtex4-2}
\usepackage{natbib}
\usepackage{graphicx}
\usepackage{bm}
\usepackage{amssymb}
\usepackage{color}
\usepackage{amsmath}
\usepackage{mathrsfs} 
\usepackage{amstext}
\usepackage[english]{babel}
\usepackage{latexsym}
\usepackage{array}             
\usepackage{multirow}         
\usepackage[usenames,dvipsnames]{xcolor}
\usepackage[colorlinks=true,citecolor=Blue,linkcolor=RubineRed,urlcolor=Blue]{hyperref}
\usepackage{tocloft}

\newcommand{\beq}{\begin{equation}}
\newcommand{\eeq}{\end{equation}}
\newcommand{\beqa}{\begin{eqnarray}}
\newcommand{\eeqa}{\end{eqnarray}}

\begin{document}

\title{Kibble-Zurek Mechanism and Beyond: Lessons from  a Holographic Superfluid Disk}

\author{Chuan-Yin Xia}
\affiliation{Center for Gravitation and Cosmology, College of Physical Science and Technology, Yangzhou University, Yangzhou 225009, China}

\author{Hua-Bi Zeng}
\email{zenghuabi@hainanu.edu.cn}
\affiliation{Center for Theoretical Physics, Hainan University, Haikou 570228, China}
\affiliation{Center for Gravitation and Cosmology, College of Physical Science and Technology, Yangzhou University, Yangzhou 225009, China}
\author{Andr\'as Grabarits\href{https://orcid.org/0000-0002-0633-7195}{\includegraphics[scale=0.05]{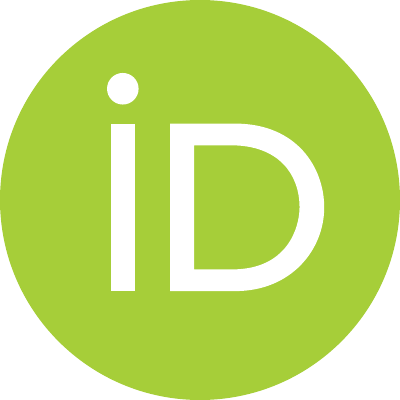}}}
\address{Department of Physics and Materials Science, University of Luxembourg,
L-1511 Luxembourg, Luxembourg}
\author{Adolfo del Campo\href{https://orcid.org/0000-0003-2219-2851}{\includegraphics[scale=0.05]{orcidid.pdf}}}
\email{adolfo.delcampo@uni.lu}
\address{Department of Physics and Materials Science, University of Luxembourg,
L-1511 Luxembourg, Luxembourg}
\address{Donostia International Physics Center, E-20018 San Sebasti\'an, Spain}

\begin{abstract}
The superfluid phase transition dynamics and associated spontaneous vortex formation with the crossing of the critical temperature in a disk geometry 
is studied in the framework of the $AdS/CFT$ correspondence by solving the Einstein-Abelian-Higgs model in an $AdS_4$ black hole. For a slow quench, the vortex density admits a universal scaling law with the cooling rate as predicted by the Kibble-Zurek mechanism (KZM), while for fast quenches, the density shows a universal scaling behavior as a function of the final temperature, that lies beyond the KZM prediction. The vortex number distribution in both the power-law and saturation regimes can be approximated by a normal distribution. However, the study of the universal scaling of the cumulants reveals non-normal features and indicates that vortex statistics in the newborn superfluid is best described by the Poisson binomial distribution, previously predicted in the KZM regime [Phys. Rev. Lett. 124, 240602 (2020)]. This is confirmed by studying the cumulant scalings as a function of the quench time and the quench depth.  Our work supports the existence of a universal defect number distribution that accommodates the KZM scaling, its breakdown at fast quenches, and the additional universal scaling laws as a function of the final value of the control parameter.

\end{abstract}

\maketitle

\section{Motivation}
The description of strong coupling problems in quantum many-body systems is challenging in equilibrium and far from it. In this context, the use of gauge-gravity duality, AdS/CFT duality, or holography offers an exciting prospect by relating a quantum field theory without gravity to a classical gravity with one more dimension  \cite{Maldacena1999,Gubser1998,witten1998}. 
Strongly coupled systems can be calculated non-perturbatively and accurately within the framework of weak-coupling gravity theory, and the use of holographic techniques has already proved useful in condensed matter theory (AdS/CMT)\cite{zaanen2015,ammon2015,hartnoll2018,zaanen2021,Liu2020}.

An important application of the holographic setup is the description of strongly-coupled systems with broken continuous symmetries such as a superconductor \cite{Hartnoll2008,Gubser2008,Herzog2009}.
Such formulation not only sheds insights into the equilibrium state of superconductors and superfluids but also in the study of vortex dynamics \cite{Xia2019,Su2023,xia2022Abrikosov,Yang2023,Natsuume2017,Lixin2020,Wittmer2021,LanShanquan2023}, and more generally, in the tenets of nonequilibrium statistical mechanics, such as the Kibble-Zurek mechanism (KZM) 
\cite{Chesler2015,Sonner2015,Zeng2019,xia2020,Li:2022tab,Li2020,Li2021,Li2022,Li2022d}.
The KZM was initially proposed in cosmology and then applied in other fields, ranging from condensed matter physics to quantum computing \cite{Kibble1976,Kibble1980,Zurek1985,Zurek1996,Dziarmaga05,Polkovnikov2011,delCampo2014}. It links the dynamics of continuous phase transitions to the equilibrium critical exponent, explains the generation of topological defects, and predicts that their density follows a universal power law that experiments and simulations can test.
According to KZM, when a system is quenched by modulating a control parameter $T$ across a critical point $T_c$ in a finite quench time scale $\tau_Q$, topological defects form as a result of the breaking of adiabaticity. The latter is unavoidable when crossing a continuous phase transition in the thermodynamics limit due to the divergence of relaxation time, which is responsible for the critical slowing down. Specifically, KZM exploits the scaling relations for the relaxation time $\tau$ and the correlation length $\xi$, encoded in the power laws $\tau=\tau_{0}|\epsilon|^{-z\nu}$  and $\xi=\xi_{0}|\epsilon|^{-\nu}$, respectively.  Here, $\nu$ is the correlation-length critical exponent, and $z$ is known as the dynamic critical exponent. The parameters $\tau_0$ and $\xi_0$ are microscopic constants, and $\epsilon$ determines the  proximity to the critical point, being defined as
\begin{equation}
\epsilon(T)=\frac{(T_c-T)}{T_c}.
\end{equation}
Consider the modulation in time of the control parameter $T(t)$, and thus, of $\epsilon(t)=\epsilon[T(t)]$. 
According to the adiabatic-impulse approximation, in the high-symmetry phase, the system adapts adiabatically to the modulation of the control parameter $\epsilon(T)$ until the time left until reaching the critical point is comparable with the instantaneous relaxation time. In the subsequent stage of the quench, macroscopic degrees of freedom are then effectively frozen until, once in the broken-symmetry phase, the order parameter starts to grow in a time scale set by the freeze-out time. 
The latter is identified by matching the instantaneous relaxation time is equal to the so-called freeze-out time.
For a linear quench near the critical point,   
\begin{eqnarray}
\label{linearquench}
T(t)=
\begin{cases}
T_c(1-t/\tau_Q)  & 0 \leq t<t_f \\
T_f              & t \geq t_f
\end{cases}.
\end{eqnarray}
The time $t_f$ is the time when the quench is completed, satisfying $T(t_f)=T_f$ that is, $t_f=\tau_Q(1-T_f/T_c)$.
The freeze-out timescale, in this case, reads
\begin{equation}
\hat{t}=(\tau_0 \tau_{Q}^{z\nu})^{\frac{1}{1+z\nu}}.\label{tfreeze1}
\end{equation}
KZM estimates the average domain size after the finite-time crossing of the phase transition by the value of the equilibrium correlation length  $\hat{\xi}$ 
 at  the freeze-out time $\hat{t}$, i.e.,  $\hat{\xi}\propto\xi(T(\hat{t}))=\xi_0(\tau_{Q}/\tau_0)^{\frac{ \nu}{1+z\nu}}$.
 Topological defects are formed with a certain probability at the junction between adjacent domains. 
As a result, the average defect number $n$ inherits a power-law dependence on the quenching time $\tau_Q$ 
\begin{equation}
n  \propto \frac{1}{\hat{\xi}^d} \propto\tau_{Q}^{-\frac{ d \nu}{1+z\nu}},\label{kzlaw}
\end{equation}
where $d$ is the spatial dimension. This is the central prediction of the KZM. Its verification has motivated continuous research efforts over decades. Early experiments focused on probing the spontaneous formation of defects in several systems, such as liquid crystals \cite{Chuang1991,Bowick1994,Digal1999}, and superfluid helium \cite{Hendry1994,Bauerle1996,Ruutu1996,Dodd1998}. Subsequent efforts focused on verifying the power-law scaling, using Josephson junctions \cite{Carmi2000,Monaco2002,Monaco2003,Monaco2006}, thin film superconductors \cite{Maniv2003,Golubchik2010}, linear optical quantum simulators \cite{Xiao-Ye2014}, trapped ions \cite{Pyka2013,Ulm2013,Ejtemaee2013}, quantum annealers \cite{Bando2020,King2022} and ultra-cold gases \cite{Sadler2006,Weiler2008,Lamporesi2013,Navon2015,Donadello2016,Ko2019}. 

This work explores the universal character of the critical dynamics beyond the KZM using a holographic superfluid phase transition as a test bed. It establishes that the mean density, considered by the KZM, and the complete defect number distribution are universal. Such universality is not restricted to slow quenches with low defect density where the KZM applies but extends to arbitrary values of the quench time. In particular, it holds even for sudden quenches, for which the defect density saturates at a plateau value, leading to a universal breakdown of the KZM.
Our work specifically demonstrates that the defect number distribution is described by a Poisson binomial distribution and that its cumulants exhibit power-law scaling laws as a function of the quench time and the depth of the quench. These predictions are experimentally testable by probing the spontaneous vortex formation scenarios in newborn superfluids.

The following section details the relevant background on physics beyond KZM, deviations from KZM, and the role of boundary conditions. A comparison of our findings with previous literature is provided in Sec. \ref{SecLit}.   
Section \ref{SecHolo} contains the holographic setting, which can be skipped by readers solely interested in the universal aspects of critical dynamics.  The universal dynamics for slow and fast quenches is established in Sec. \ref{SecFastSlow}. The discussion \ref{SecDiss} establishes the Poisson binomial distribution as the universal distribution that describes defect statistics as a function of the quench time and depth in and out of the KZM regime. We close with a summary and outlook in Sec. \ref{SecSumm}.

\section{Background}
\subsection{Beyond KZM physics}
Beyond the KZM predictions, the fluctuations of the number of topological defects have been predicted to be universal. In one-dimensional systems, the point-like defect number distribution is well described by can usually be described by a binomial or Poisson-binomial distribution \cite{delCampo2018,GomezRuiz2020}. As a result, the universal scaling with the quench time extends beyond the KZM prediction to the variance and any other higher-order cumulant \cite{delCampo2018,GomezRuiz2020}, as confirmed by numerical simulations \cite{delCampo2018,GomezRuiz2020,Mayo2021,delCampo2021,Subires2022,Li2022,Li2022d,Li:2022tab} and experiments \cite{Cui2020,Bando2020R,Goo2021,King2022}. KZM and beyond-KZM physics have also been established by strongly-coupled systems described by holography \cite{Chesler2015,Natsuume2017,Zeng2019,Zeng2023,xia2020,Xia2021xap,Xia2023,Li:2022tab,Li2020,Li2021,Li2022,Li2022d,Sonner2015}, showcasing their universality.  
Specifically, numerical simulations have been reported for one-dimensional \cite{Sonner2015,xia2020,Li:2022tab,Li2022,Li2022d}  and two-dimensional \cite{Chesler2015,Li2020,Li2021} holographic superconductors and superfluids. But for some features \cite{Chesler2015,Sonner2015}, all results are rather similar to those of the mean-field theory and verify the validity of KZM in holographic systems.

\subsection{Deviations from KZM physics and fast-quench universality}
Naturally, the validity of the KZM scaling laws is limited by the onset of adiabaticity that occurs when $\hat{\xi}$ is comparable to the system size. 
There are also reports on the impact of finite size on KZM \cite{delCampo2010,Corman2014,xia2020}. When the system size is close to the correlation length, the number of topological defects will deviate from the prediction of KZM. Numerical studies of the finite-size effects  \cite{xia2020} show deviations of KZM universality, explaining experimental findings in Ref. \cite{Corman2014}.

By contrast, when the quenching time is much shorter than the time scale in which the order parameter grows,  the density of topological defects no longer satisfies the universal power law predicted by the KZM. Instead, it saturates at a constant value called the plateau. This phenomenon was found in experiments \cite{Donadello2016,Ko2019,Goo2021,Goo2022} after being observed in numerical simulations \cite{delCampo2010,Chiara2010,Sonner2015,GomezRuiz2019,xia2020,Xia2021xap,Zeng2023,Xia2023,Xia2021xap,Zeng2019,Li2020,Li2021}. This universal breakdown of the KZM scaling at fast quenches is to be expected, as its derivation is restricted to slow quenches near the adiabatic regime \cite{Dziarmaga05,Polkovnikov2011,Grandi2010,delCampo2014}.  
The nature of the plateau, the value of the plateau density, and its onset have remained a matter of debate.  Recently \cite{Zeng2023}, its universal character was established, along with the fact that the plateau density is independent of the quench rate, being governed by the equilibrium relaxation time associated with the final value of the control parameters. These findings have been further verified in a holographic setting involving a structural phase transition \cite{Xia2023}.

For rapid quenches, the end time $t_f$ is smaller than the relaxation time, that is,
\begin{equation}
t_f\leq \tau(t_f)=\tau(T_f), \label{rapidconditon}
\end{equation}
so the freezing will not occur when $0<t<t_f$. As the time of evolution goes by and  $t$ increases, it eventually matches the minimum relaxation time $\tau(T_f)$. Thus, the freeze-out time in a rapid quench scales universally as a function of   $\epsilon_f=(T_c-T_f)/T_c$
\begin{equation}
\hat{t}=\tau(T_f) \propto \epsilon_{f}^{-\nu z}, \label{rapidtf}
\end{equation}
at variance with the KZM prediction (\ref{tfreeze1}). In addition, the correlation length is also determined by the final coupling value $T_f$ at the freeze-out time, leading to  the average number of defects 
\begin{equation}
n\propto \frac{1}{\xi^d(T_f)} \propto \epsilon_{f}^{d \nu}. \label{rapidn}
\end{equation}
To sum up, quench protocols that satisfy the constraints of Eq. \eqref{rapidconditon} can be called rapid. They lead to a non-equilibrium configuration of the system that is solely determined by the final value of the coupling constant $T_f$ and the equilibrium critical exponents. As a result, the defect density Eq. \eqref{rapidn} forms a plateau in the rapid quenching region, independent of the quench time. 
Furthermore, by matching the quench end time $t_f$ and the relaxation time $\tau(T_f)$, the crossover quenching time $\tau_{Q}^{c1}$ can be obtained, as shown in Fig. \ref{fig2KZlaw} below. It is called the first critical quench rate and satisfies the following condition
\begin{equation}
t_f=\tau_{Q}^{c1}(1-T_f/T_c)=\tau(T_f)=\tau_{0}(T_c-T_f)^{- \nu z},
\end{equation}
which yields 
\begin{equation}
\tau_{Q}^{c1}\propto\epsilon_{f}^{-(\nu z+1)}.\label{tauQc1}
\end{equation}
These results establish the universality of critical dynamics in rapid quenches in a holographic framework.

\subsection{Role of boundary conditions}
In addition to these advances, the possible breakdown of KZM and beyond-KZM predictions has been explored in various scenarios. This includes the effect of boundary conditions on KZM. While open boundary conditions are natural in most experiments, numerical simulations are often facilitated by imposing periodic boundaries. It is further known that the role of boundary conditions cannot be ignored, regardless of the system size. For instance, the Poincar\'e-Hopf theorem dictates that the total topological charge of spontaneously formed defects needs to vanish identically in manifolds with zero Euler characteristic \cite{Musevic17}. As a result, breaking of $\mathbb{Z}_2$ symmetry restricts defect formation to kink-antikink pairs on a ring topology \cite{GomezRuiz2022}. Similarly, the thermal quench of a superconducting or superfluid ring leads to a vanishing winding number \cite{Zurek1996,Dziarmaga08,Das12,Sonner2015,Nigmatullin16,xia2020}. 
In the same manner, breaking of $U(1)$ associated with vortex formation is conditioned to vanishing total vorticity in samples with a torus topology \cite{delCampo2021,Thudiyangal24}. 
Results obtained in a lattice model indicate that 
 even under extremely slow quenching, periodic, anti-periodic, and free boundary conditions will not suppress the universal power law scaling for the mean density but may affect high-order cumulants and the defect number distribution \cite{GomezRuiz2022}. 
In studies of KZM physics, experimental setups often involved open boundaries  \cite{Hendry1994,Bauerle1996,Ruutu1996,Dodd1998,Maniv2003,Golubchik2010,Sadler2006,Weiler2008,Lamporesi2013,Navon2015,Donadello2016,Ko2019}. However, two-dimensional systems with boundaries have not been numerically simulated in a holographic setting.
In this article, we use the holographic superfluid model on an open disk geometry \cite{hartnoll2018} as a test bed for vortex formation and the validity of both KZM and beyond-KZM physics. As we use Neumann boundary conditions, the Poincar\'e-Hopf theorem does not restrict the total vorticity. This geometry involves a boundary at the maximum radius, making it possible to test whether the finite-size boundary affects the universal finite-quench scaling behavior.

\section{Findings and comparison to previous works}\label{SecLit}

In this work, we demonstrate the far-from-equilibrium universality in the defect number distribution both as a function of the quench time and the quench depth, that is, the final value of the control parameter.
Our work largely transcends the scope of the celebrated KZM, focused on the mean density of topological defects as a function of the quench time \cite{delCampo2014}. Likewise, our findings extend the notion of the universality of critical dynamics with respect to previous works on beyond-KZM physics.
Specifically, previous works reporting the universality of the defect number statistics in classical and quantum systems are restricted to the power-law scaling regime associated with slow quenches in which KZM holds, in theoretical  \cite{delCampo2018,GomezRuiz2020,Mayo2021,delCampo2021,Subires2022,Li2022,Li2022d,Li:2022tab} and experimental studies \cite{Cui2020,Bando2020R,Goo2021,King2022}. Our manuscript established the universality of the defect number distribution at arbitrarily quench rates for the first time, from the slow to the sudden regimes. Specifically, we establish that the Poisson binomial distribution describes the universal statistics of spontaneously formed point-like topological defects in all regimes. 
It is a curious accident that previous studies have evaded the universal signatures of the defect number statistics in the power-law scaling regime for decades, possibly due to the lack of a theoretical framework, despite the signal being in the noise often measured in the experiments. Once noticed, such a universal distribution is compatible with the observation that the KZM regime is restricted to slow quenches with very low defect densities. Our results are remarkable in that they establish the universality of the defect number distribution even at the highest density values achievable by spontaneous defect formation through sudden quenches that are associated with the saturation plateau and the breakdown of the KZM.

\section{Model of  holography superfluid disk and numerical scheme for phase transition dynamics}\label{SecHolo}
A simple action for a holographic superfluid  consists of a complex scalar field $\Psi$ with mass $m$, minimally coupled to a $U(1)$ gauge field $A_\mu$ \cite{Gubser2008,Hartnoll2008,Herzog2009}, 
\begin{equation}
S=\int  d^4x \sqrt{-g}\Big[-\frac{1}{4}F^2-|D\Psi|^2-m^2|\Psi|^2\Big],
\label{model}
\end{equation}
where $F_{\mu\nu}=\partial_\mu A_\nu-\partial_\nu A_\mu$, $D_\mu=\partial_\mu-iqA_\mu$ with charge $q$.  The theory can be defined in a $AdS_4$ black hole background with Eddington-Finkelstein coordinates,
\begin{equation}
ds^2=\frac{\ell^2}{u^2}\left(-f(u)dt^2-2dtdu + dr^2+ r^2d\theta^2\right),
\label{metric}
\end{equation}
in which $\ell$ is the AdS radius, $u$ is the AdS radial coordinate of the bulk
and $f(u)=1-(u/u_h)^3$.  Thus, $u=0$ is the AdS boundary, while $u=u_h$ is the horizon; $r$ and $\theta$  are, respectively, the radial and angular coordinates of the dual $2+1$ dimensional boundary, which is a holographic superfluid disk as in Ref. \cite{Xia2019,Su2023}. The Hawking temperature is $T=3/(4\pi u_h)$. From dimensional analysis, the ratio $T/\mu$ between the Hawking temperature $T$ and the chemical potential $\mu=A_t|_{u=0}$  is dimensionless. The temperature on the boundary can be expressed as
\begin{equation}
\label{Temperature}
T=T_c\frac{\mu_c}{\mu}.
\end{equation}
Thus, increasing the charge density is equivalent to decreasing the temperature.
For simplicity, the probe limit is adopted in this work by assuming that the matter fields do not affect the gravitational fields. After rescaling $\ell = u_h = 1$,  the equations of motions (EoMs) can be written as
\begin{equation}
(-D^2+m^2)\Psi=0,~~~  \partial_\mu F^{\mu\nu}=J^\nu,
\label{eom}
\end{equation}
where $J^{\mu}=i (\Psi^*D^{\mu} \Psi-\Psi D^{\mu} \Psi^*)$
is the bulk current. 
Furthermore, the axial gauge $A_u=0$ is adopted as in \cite{Herzog2009}. 
Without loss of generality, we set $m^2$ to $-2$.

The fully expanded equation of motion can be written as
\begin{equation}
\begin{split}
-\left(\partial_{u}\left(f\partial_{u}\Psi\right)+i\left(\partial_{u}A_{t}\right)\Psi+2iA_{t}\partial_{u}\Psi\right)\\
+\left(-\partial_{r}^{2}\Psi+i\left(\partial_{r}A_{r}\right)\Psi+2iA_{r}\partial_{r}\Psi-\frac{1}{r}\partial_{r}\Psi\right)\\
+\frac{1}{r^{2}}\left(-\partial_{\theta}^{2}\Psi+i\left(\partial_{\theta}A_{\theta}\right)\Psi+2iA_{\theta}\partial_{\theta}\Psi\right)\\
+\left(A_{r}^{2}+\frac{iA_{r}}{r}+\frac{A_{\theta}^{2}}{r^{2}}+u\right)\Psi+2\partial_{t}\partial_{u}\Psi =0,
\label{eom1}
\end{split}
\end{equation}

\begin{equation}
\begin{split}
\partial_{u}^{2}A_{t}-\partial_{u}\partial_{r}A_{r}-\frac{1}{r}\partial_{u}A_{r}-\frac{1}{r^{2}}\partial_{u}\partial_{\theta}A_{\theta}\\
-i\left(\Psi^{*}\partial_{u}\Psi-\Psi\partial_{u}\Psi^{*}\right)=0,
\label{eom2}
\end{split}
\end{equation}

\begin{equation}
\begin{split}
-\frac{1}{r}\partial_{u}A_{r}-\partial_{t}\partial_{u}A_{t}-\partial_{t}\partial_{r}A_{r}-\frac{1}{r^{2}}\partial_{t}\partial_{\theta}A_{\theta}+\frac{f}{r}\partial_{u}A_{r}\\
+f\partial_{u}\partial_{r}A_{r}+\frac{f}{r^{2}}\partial_{u}\partial_{\theta}A_{\theta}+\partial_{r}^{2}A_{t}+\frac{1}{r}\partial_{r}A_{t}+\frac{1}{r^{2}}\partial_{\theta}^{2}A_{t}\\
-i\left(\Psi^{*}\partial_{t}\Psi-\Psi\partial_{t}\Psi^{*}\right)-2A_{t}\Psi^{*}\Psi\\
+if\left(\Psi^{*}\partial_{u}\Psi-\Psi\partial_{u}\Psi^{*}\right)=0,
\label{eom3}
\end{split}
\end{equation}

\begin{equation}
\begin{split}
2\partial_{t}\partial_{u}A_{r}-\partial_{u}\partial_{r}A_{t}-\partial_{u}\left(f\partial_{u}A_{r}\right)+\frac{1}{r^{2}}\left(\partial_{r}\partial_{\theta}A_{\theta}
-\partial_{\theta}^{2}A_{r}\right)\\
+i\left(\Psi^{*}\partial_{r}\Psi-\Psi\partial_{r}\Psi^{*}\right)+2A_{r}\Psi^{*}\Psi=0,
\label{eom4}
\end{split}
\end{equation}

\begin{equation}
\begin{split}
2\partial_{t}\partial_{u}A_{\theta}-\partial_{u}\left(f\partial_{u}A_{\theta}\right)-\partial_{u}\partial_{\theta}A_{t}-\partial_{r}^{2}A_{\theta}+\frac{1}{r}\partial_{r}A_{\theta}\\
+\partial_{r}\partial_{\theta}A_{r}-\frac{1}{r}\partial_{\theta}A_{r}+i\left(\Psi^{*}\partial_{\theta}\Psi-\Psi\partial_{\theta}\Psi^{*}\right)+2A_{\theta}\Psi^{*}\Psi=0.
\label{eom5}
\end{split}
\end{equation}
The five partial differential equations 
Eq. \eqref{eom1}-\eqref{eom5} are not independent, so we can choose any four of them. In this work, we  choose Eq. \eqref{eom1}-\eqref{eom4} as in Ref. \cite{Xia2019}.
In addition to the above settings, proper boundary conditions should be imposed to solve them.
We impose the regular boundary conditions of all physical solutions
at the horizon of the black hole. Explicitly, we set $A_t=0$ on the horizon as in \cite{Natsuume2017,Xia2019,Su2023}. Boundary conditions for other fields at the horizon are automatically satisfied because they are implicit in their respective equations rather than being set on an ad hoc basis.

Near the boundary $u=0$, the general solutions take the asymptotic form as,
\begin{eqnarray}
A_\nu(t,u,r,\theta)&=& a_\nu(t,r,
\theta)+ b_\nu(t,r,\theta) u+\mathcal{O}(u^2),
\label{aboundary} \\
 \Psi(t,u,r,\theta)&=& \Psi_1(t,r,\theta) u+ \Psi_2(t,r,\theta) u^2+\mathcal{O}(u^3).
 \label{psiboundary}
\end{eqnarray}
From the AdS/CFT dictionary, the coefficients $a_{r,\theta}$ can be treated as the superfluid velocity along the coordinates $r, \theta$, while $b_{r,\theta}$ are the conjugate currents \cite{Montull2009}.  The coefficients $a_t$ and $b_t$ are respectively interpreted as the chemical potential and charge density in the boundary field theory; $\Psi_1$ is the source term, while $\Psi_2$ is the vacuum expectation value $\langle O\rangle$ of the dual scalar operator.
On the boundary $u=0$, we always set $\Psi_1\equiv0$ in order to have spontaneously broken symmetry of the order parameter. We also fix the chemical potential, that is, $a_t=\mu$.
From Eq. \eqref{Temperature}, when $\mu>\mu_c$, the temperature $T$ below $T_c$, we can get the superfluid phase.  The critical chemical potential in our setting takes the value $\mu_c\sim4.0637$. The critical exponents of the phase transition of this model are consistent with those of the mean-field theory \cite{Maeda2009,Jensen2011}. 
\begin{figure}
    \centering
    \includegraphics[width=1\linewidth]{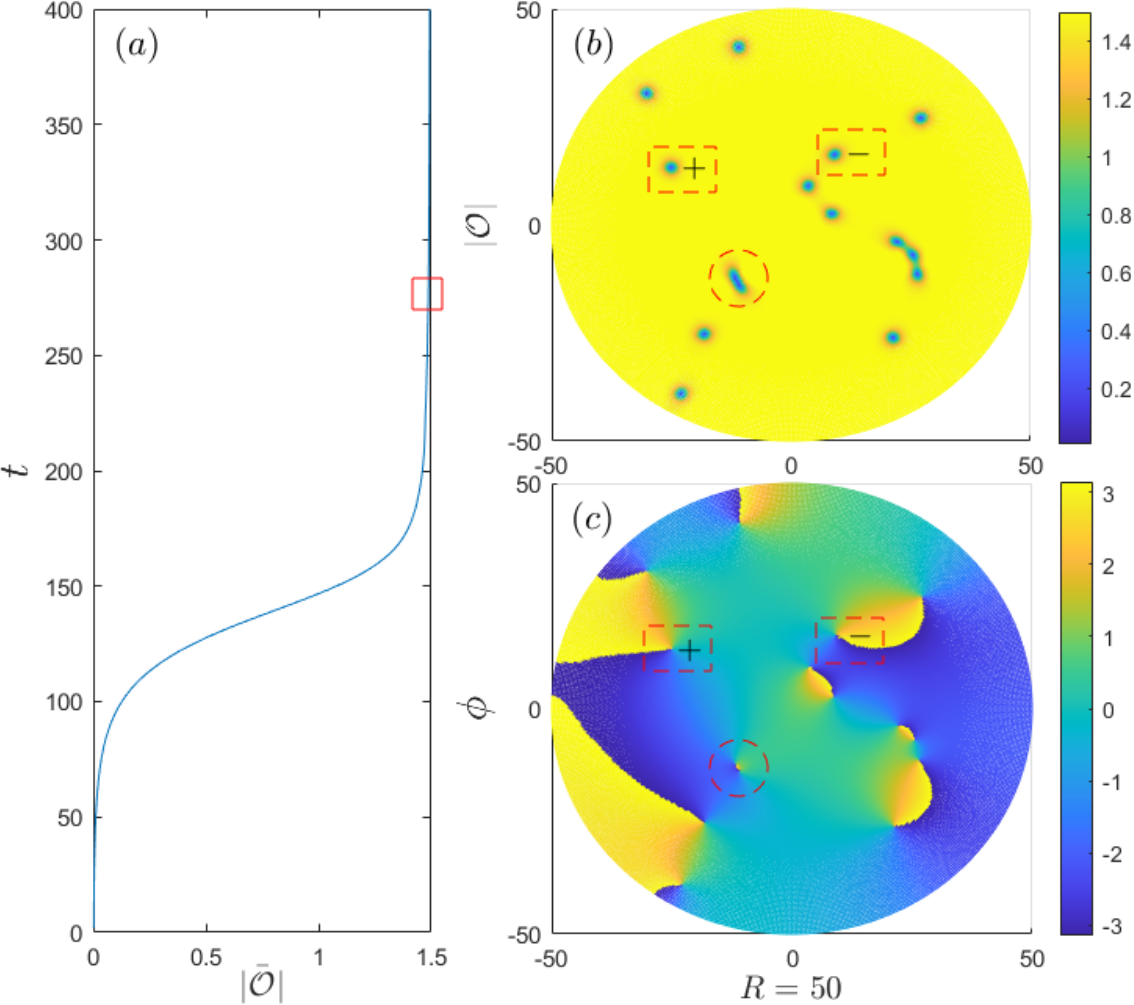}
    \caption{ Vortices and anti-vortices formed in a holographic superfluid disk after thermal quench: (a)  Growth of the average order parameter as a function of time, where $\tau_Q=10$, $T_f=0.94T_c$ and $R=50$. Panels (b) and (c) correspond to snapshots of the spatial dependence of absolute value and phase of the local order parameter at the quasi-equilibrium stage ($t=275$) of the phase transition marked by red squares in panel (a). A red rectangular dashed line surrounds a positive vortex and an antivortex, while a red circular dashed line surrounds a vortex pair. The snapshot has 8 positive vortices and 7 anti-vortices, among which the vortex pairs are annihilating. }
    \label{fig1Order_phase}
\end{figure}

There are two spatial coordinates 
 $\theta,r$  whose boundary conditions need to be specified. For a disk, all fields along the $\theta$ have periodic boundary conditions. 
When a phase transition occurs, the condensed value of the order parameter grows dynamically, so we cannot fix it.
Along the radial coordinate of the disk, we impose  Neumann boundary conditions $\partial_r A_{\nu}=\partial_r\Psi=0$ at $r=R$, rather than Dirichlet boundary conditions. In our model, $R$ is the radius of the disk embedded in the boundary of the bulk. Note that this Neumann boundary condition is imposed on the whole range of $u$.

Chebyshev spectral methods are used in the $(u,r)$ direction for numerical simulation. 
Fourier spectral methods are used in the $\theta$ direction. 
In the time direction, we use the fourth-order Runge-Kutta method to simulate the evolution of the system and use boundary conditions at each step of the evolution to constrain the values of the boundaries in other directions.
In addition, all fields need to be given initial values. The specific scheme will be introduced in detail in the next section.

Before driving the system by a thermal quench, the normal fluid should be prepared above the critical temperature at equilibrium.
Numerically, we first give a set of normal state solutions at $\mu=\mu_c$ corresponding to the critical temperature $T_c$.  It is uniform in the $r$ and $\theta$ direction,  and written as
\begin{eqnarray}
& A_t =\mu_c(1-u),\nonumber\\
& A_x=A_y=\Psi=0.
\end{eqnarray}
Then, we added random noise to $\Psi$ in the hole bulk to simulate the thermal fluctuation of the system, that is, $\Psi=S(t,r,\theta)$.
In this way, the normal fluid that grows at the boundary also inherits random noise $s(t,r,\theta)$. The noise $s$ satisfies the normal distribution, with a vanishing statistical average $\langle s(t,r,\theta) \rangle=0$, 
and two-point correlations $\langle s(t,r,\theta)s(t',r',\theta ') \rangle= h  \delta(t-t',r-r',\theta-\theta ')$, where the amplitude $h={10}^{-4}$. After a little heating, the fluid disk is ready for 
quenches.

\section{Universal scaling in slow and fast quench region}\label{SecFastSlow}

Using Eq. \eqref{Temperature}, we can simulate the linear cooling process by varying the chemical potential in the following way
\begin{eqnarray}
\mu(t)=
\begin{cases}
\mu_c(1-t/\tau_Q)^{-1}  & 0 \leq t<t_f \\
\mu_f              & t \geq t_f
\end{cases}.
\end{eqnarray}
The time $t_f$ at which the modulated coupling reaches its
final value, $\mu(t_f)=\mu_f$, will play a key role in what follows
and is given by
\begin{equation}
t_f=\tau_Q(1-\mu_c/\mu_f).
\end{equation}

A single realization after the thermal quench is shown in Fig. \ref{fig1Order_phase} for the choice of parameters $\tau_Q=10$, $T_f=0.94T_c$, and $R=50$.  Like the superfluid in the boundaryless tread geometry \cite{Chesler2015}, the superfluid in the disk geometry with open boundaries can also generate vortices through the KZM, and the vortices will also annihilate. However, the difference is that the number of positive and negative vortices need not be equal in every quenching experiment, only on average. 
 Positive and negative vortices adjacent to each other will annihilate in pairs. This phenomenon is in stark contrast to what occurs in a rotating superfluid under disk geometry \cite{Xia2019,Su2023}, when all vortices have equal circulation. In addition, vortices may also be lost at the disk boundary.

\begin{figure}
    \centering
    \includegraphics[width=1\linewidth]{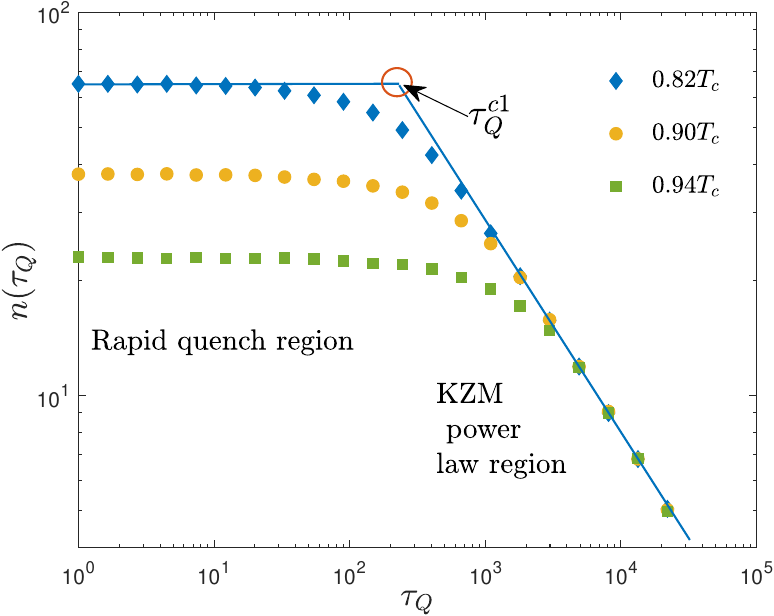}
    \caption{Universal scaling laws at fast and slow quenches in a log-log representation: The mean number $n$ of kinks as a function of the quench time $\tau_Q$. The behavior at slow quenches exhibits a universal power-law scaling with the quench time, which is in agreement with KZM. As the quench time is reduced, there is a crossover to a plateau region with a value of the vortex average number that is independent of the quench time. The crossover occurs for $\tau_Q$ shorter than $\tau_Q^{c1}$. The data in the KZM power law region is fitted by the $n=(992\pm 71)\tau_Q^{-0.518\pm 0.018}$.}
    \label{fig2KZlaw}
\end{figure}

\begin{figure}
    \centering
    \includegraphics[width=1\linewidth]{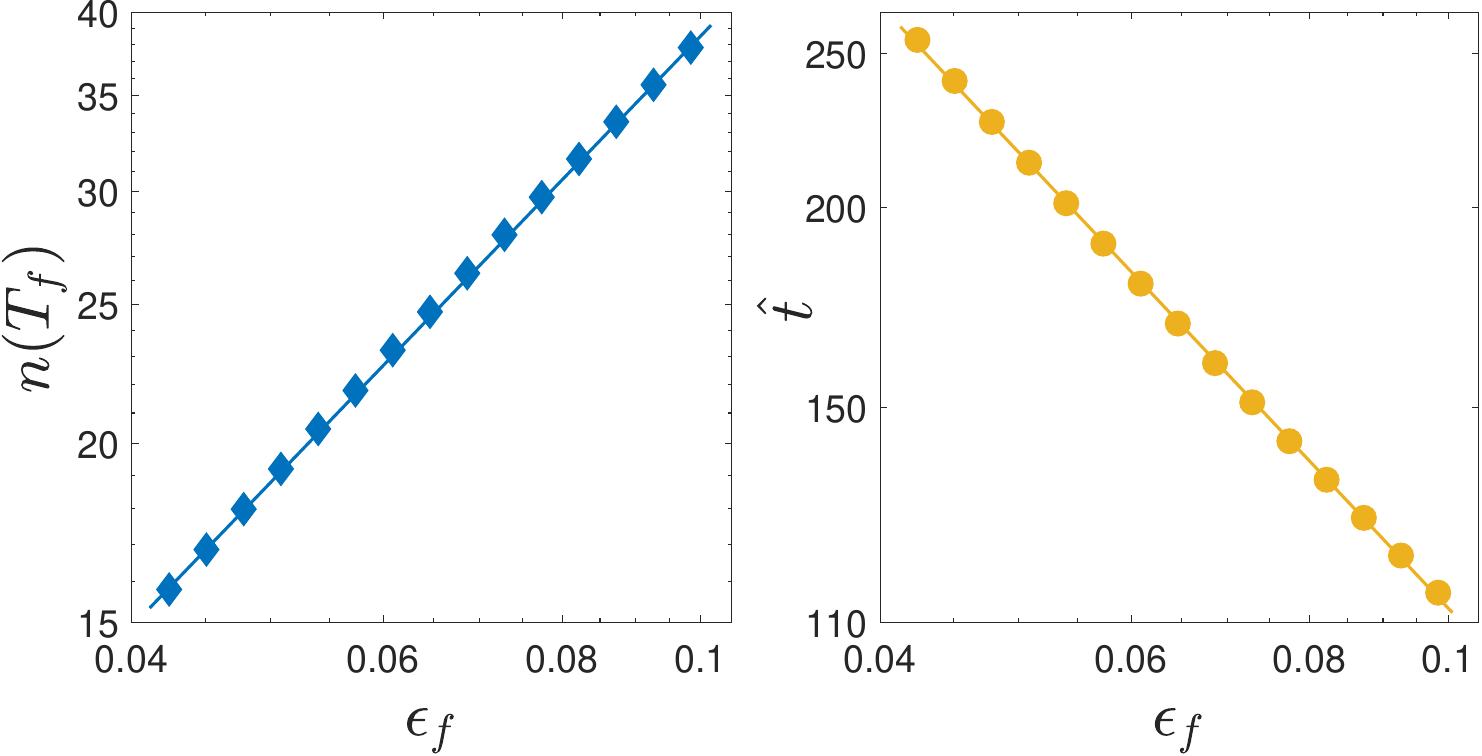}
    \caption{Universal scaling laws at fast quenches in a log-log representation. (a) The mean number $n$ of vortices and (b) the freeze-out time $\hat{t}$ scale universally as a function of the quench depth $\epsilon_{f}$. The scaling laws are determined using about $10^5$  trajectories. The corresponding data is fitted by the following power-laws  $n=(416\pm 10)\epsilon_f^{1.03\pm 0.01}$, $\hat{t}=(12.4\pm 0.2)\epsilon_f^{-0.957\pm 0.008}$.}
    \label{fig3rapidquench}
\end{figure}

Fig. \ref{fig2KZlaw} reports universal scaling laws at fast and slow quenches in a log-log representation. Each sampling point is an average of 2000 realizations.
In panel \ref{fig2KZlaw}, from top to bottom, the curves for the average number of kinks are shown for three different values of the end temperatures $T_f=0.82T_c,0.90T_c,0.94T_c$, in a larger disk $R=50$.
The right sides of the three curves all meet on a common line with a
negative slope, indicative of universal nonequilibrium behavior, and fitted to $n=(992\pm 71)\tau_Q^{-0.518\pm 0.018}$. 
It is in reasonable agreement with the universal KZM power law scaling in Eq. \eqref{kzlaw}, with a power law exponent $\nu/(1+z\nu)=1/4$ using the verified values of the equilibrium critical exponents $\nu=1/2$ and $z=2$.
The proximity of the numerical value of the fitted power-law exponent to the KZM prediction supports the validity of the mean-field description and the validity of the KZM in holography.
Additionally, the plateau and deflection on the left side of Fig. \ref{fig2KZlaw} hint at the destruction of KZM.
Fig. \ref{fig3rapidquench} further shows the breakdown of the KZM scaling law Eq. \eqref{kzlaw} at fast quenches, 
On the left of the Fig. \ref{fig3rapidquench},
the results are fitted to  $n=(416\pm 10)\epsilon_f^{1.03\pm 0.01}$, 
which exhibits a universal power-law scaling of the topological defects and the quench depth  $\epsilon_f$, this is in agreement with the universal prediction \eqref{rapidn}.
The numerical points fit by $\hat{t}=(12.4\pm 0.2)\epsilon_f^{-0.957\pm 0.008}$ in the right of Fig. \ref{fig3rapidquench} also shows
the corresponding freeze-out timescales varying universally at fast quenches according to \eqref{rapidtf} as a function of the quench depth.

\section{Universal statistics of vortices}

\begin{figure}
    \centering
    \includegraphics[width=1\linewidth]{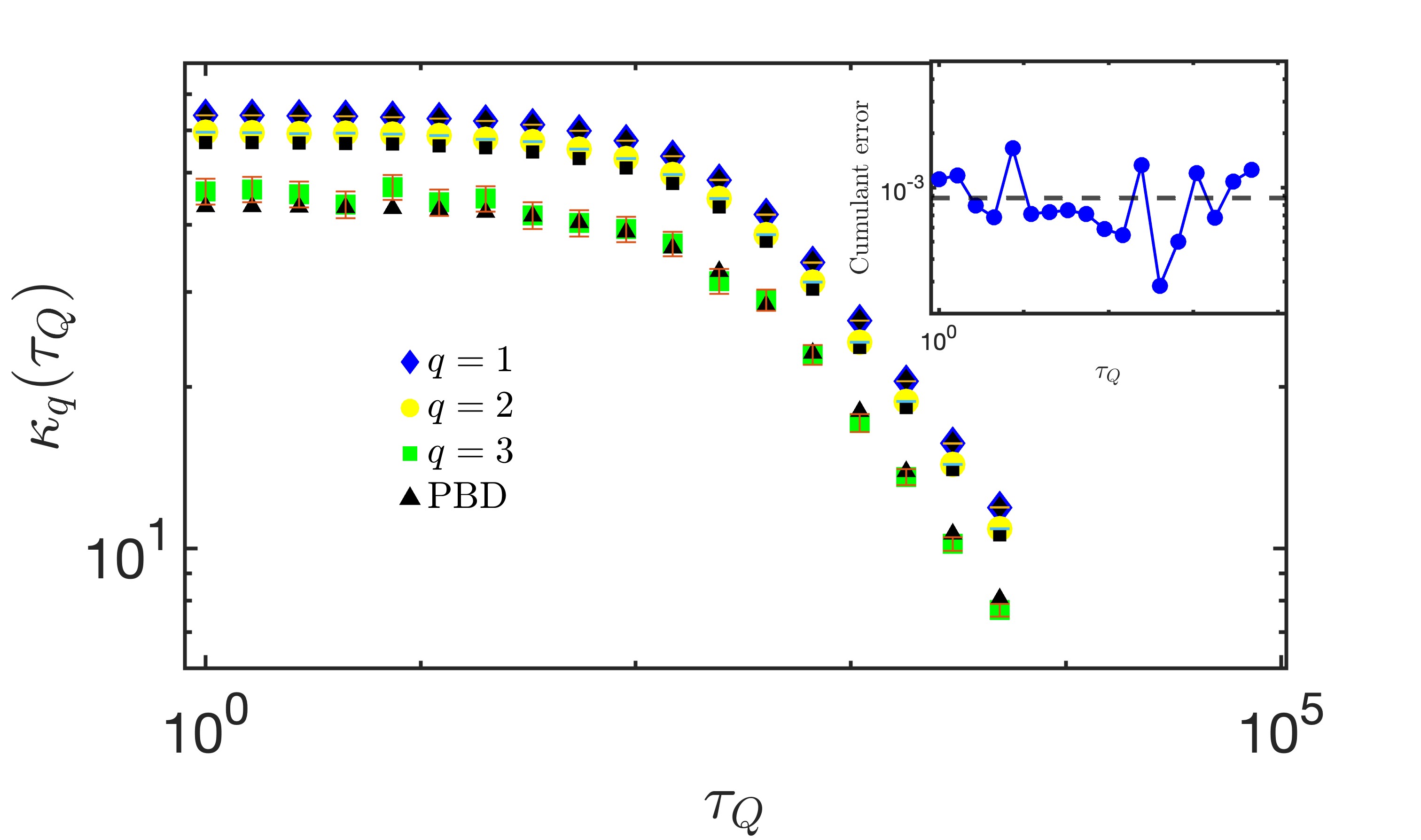}
    \caption{Universal scaling of low-order cumulants of the vortex number distribution versus the quench time $\tau_Q$. The first two cumulants $\kappa_1$ and $\kappa_2$  are the mean value and the variance of the vortex number, respectively. The third cumulant $\kappa_3$  is related to the skewness of $n$. 
    In the KZM power law region, the first cumulant $\kappa_1$ is fitted to $\kappa_1=(992\pm 71)\tau_Q^{-0.518\pm 0.018}$,  the second cumulant $\kappa_2$ is fitted to $\kappa_2=(932\pm 101)\tau_Q^{-0.522\pm0.016}$ and the third cumulant $\kappa_3$ is fitted to $\kappa_3=(670\pm 130)\tau_Q^{-0.522\pm0.029}$. For each $\tau_Q$,  $5\times 10^5$ are considered. Black triangles denote the theoretical values of a PBD of two different Bernoulli weights. The numerically fitted Bernoulli probabilities and the ratio of the average number of vortices $p_1=0.13,\,p_2=0.1,\,r=0.38$ were obtained numerically. The inset shows the relative square distance from the numerical results as a function of $\tau_Q$.}
    \label{fig4KZ}
\end{figure}

The estimation of the low-order cumulants of the vortex number distribution is based on the assumption that the generation of vortices at different locations is governed by independent and equally distributed random events \cite{GomezRuiz2020,delCampo2021,Mayo2021,Thudiyangal24}. In our model, when the normal fluid undergoes a phase transition into a superfluid, it partitioned into ``protodomains'' of size set by the correlation length $\hat{\xi}$, and there is a certain probability $p$ of vortex generation at the interface between adjacent domains. The number of possible domain locations for defect formation can be estimated as $\mathcal{N}_d=A/(f\hat{\xi}^2)$, for a newborn superfluid of area $A$, where $f$ is a fudge factor. The probability of forming $N$ defects is given by the binomial distribution 
\beqa
\label{BinomDist}
P(N)&=&B(N,\mathcal{N}_d,p)\\
&=&\binom{\mathcal{N}_d}{N}p^N(1-p)^{\mathcal{N}_d-N}. 
\eeqa
This distribution encodes the universal scaling of its cumulants $\kappa_q$ with the quench time $\tau_Q$ in the slow-quench limit
\begin{equation}\label{cumulant_Slow}
    \kappa_q\propto\left(\frac{\tau_0}{\tau_Q}\right)^\frac{2 \nu}{1+z \nu},
\end{equation}
and with the quench depth $\epsilon_f$ for fast quenches
\begin{equation}\label{cumulant_Fast}
    \kappa_q\propto\epsilon_f^\nu.
\end{equation}
Figure \ref{fig4KZ} shows the low-order cumulants of the vortex number as a function of the quench time $\tau_Q$.  A transition from the fast-quench plateau to the power-law scaling regime is observed. In the latter,  
the first three cumulants $\kappa_1$, $\kappa_2$ and $\kappa_3$ follow
the scaling law $\kappa_1=(992\pm 71)\tau_Q^{-0.518\pm 0.018}$, $\kappa_2=(932\pm 101)\tau_Q^{-0.522\pm0.016}$ and $\kappa_3=(670\pm 130)\tau_Q^{-0.522\pm0.029}$. 
The universal scaling of the cumulants generalizes the beyond-KZM predictions in \cite{GomezRuiz2020} to two-spatial dimensions, with fitted power-law exponents slightly above the expected value of $1/2$ in the mean-field regime in each case 
 ($q=1,2,3$). The situation is reminiscent of that in recent related experiments, where the observed power-law exponents for the mean slightly surpassed the theoretical prediction \cite{Navon2015,Chomaz2015,lee2023observation}. 
Beyond the mean $\kappa_1$, cumulants are highly sensitive to the quality of the sampling, and a total of half a million trajectories are used for the convergence of the third cumulant. This indicates the difficulty of probing the universal scaling of high-order cumulants. The nonzero value and scaling of the third cumulant confirm the non-Gaussian features of the underlying distribution.

\begin{figure}
    \centering
    \includegraphics[width=1\linewidth]{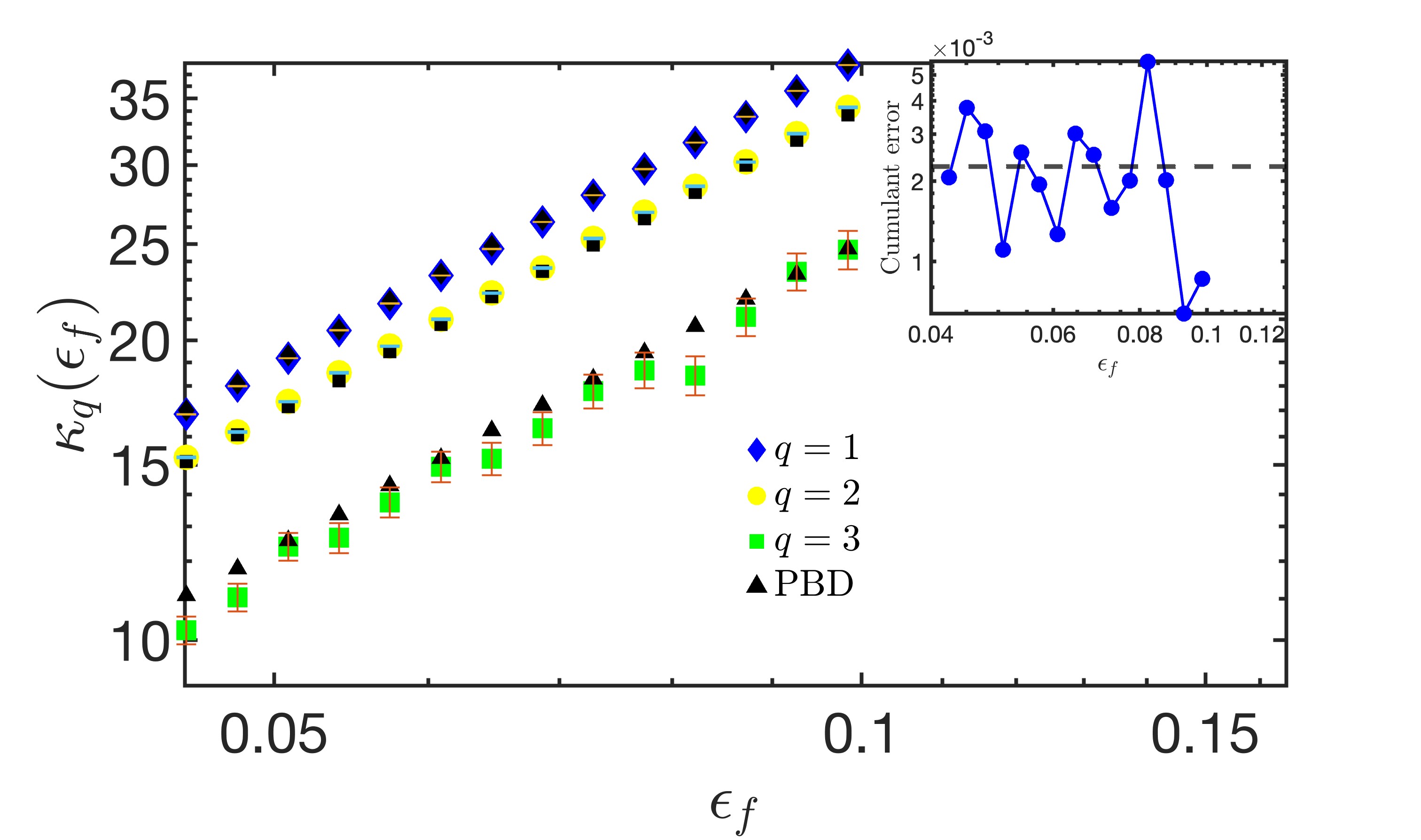}
    \caption{Universal scaling of low-order cumulants of the vortex number distribution versus the quench depth $\epsilon_f$ at fast quenches. 
     The first cumulant $\kappa_1$ is fitted to $\kappa_1=(416\pm 10)\epsilon_f^{1.03\pm 0.01}$, the second cumulant $\kappa_2$ is fitted to $\kappa_2=(378\pm 9)\epsilon_f^{1.034\pm 0.001}$ and the third cumulant $\kappa_2$ is fitted to $\kappa_3=(294\pm 9)\epsilon_f^{1.08\pm 0.08}$. Values for each $\epsilon_f$ involve $5\times 10^5$ trajectories. Black triangles denote the cumulant values obtained by the best-fitting PBD with the same parameters as in Fig.~\ref{fig4KZ}. The inset shows the relative square distance from the numerical results as a function of $\tau_Q$.}\label{fig5NEpsilon}
\end{figure}

\begin{figure}
    \centering
    \includegraphics[width=1\linewidth]{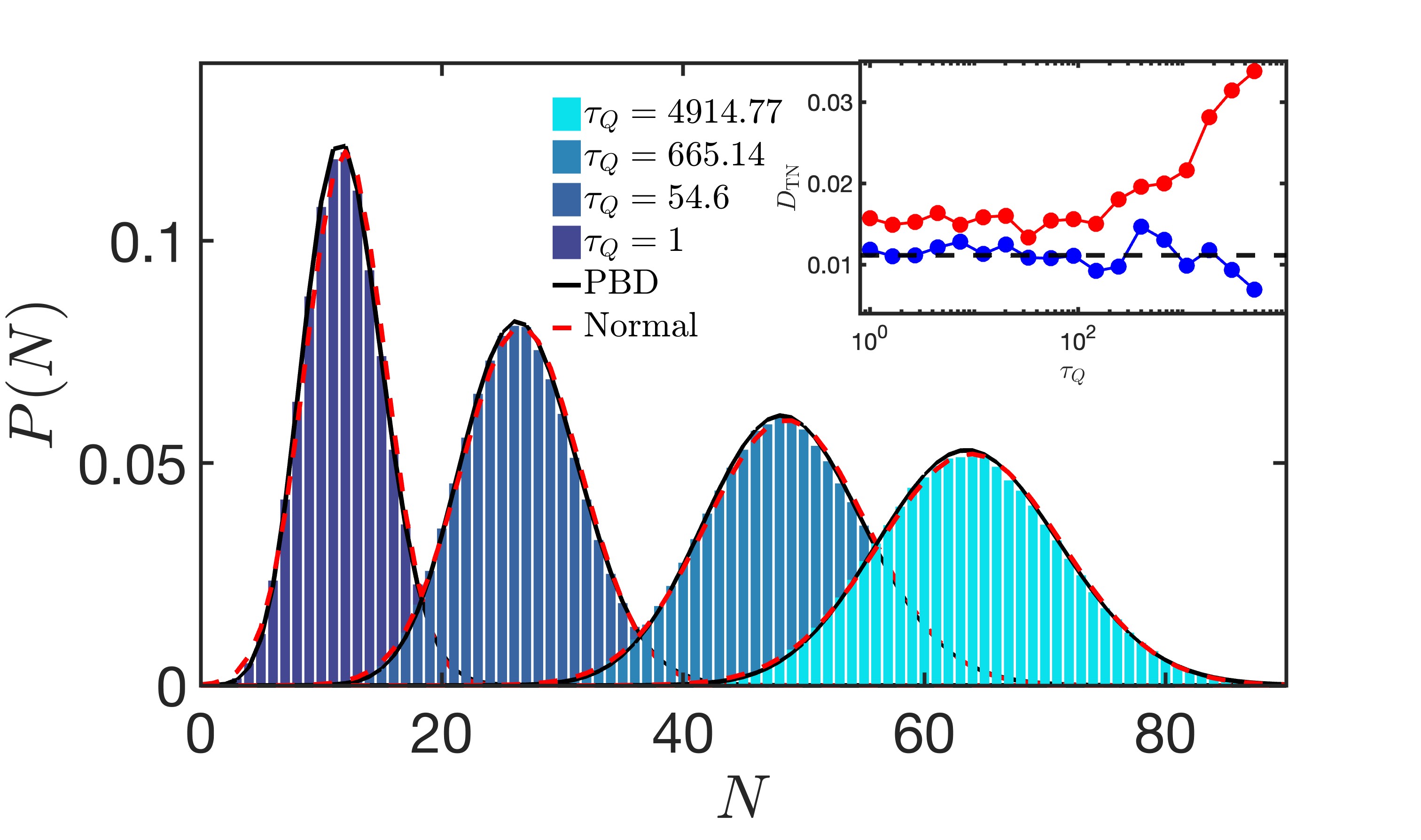}
    \caption{Histograms for the vortex-number probability distribution for different values of the quench time. 
    Both even and odd outcomes of the vortex number are possible as a result of the Neumann boundary conditions on the disk. 
    The normal approximation is shown in red. 
    The PBD  accurately fits the data in all regimes, accounting also for the non-binomial nature of the first three cumulants in Figs. \ref{fig4KZ}  and \ref{fig5NEpsilon}. The solid line denotes the PBD fitting with the same parameters as in Figs.~\ref{fig4KZ} and~\ref{fig5NEpsilon} accounting for the additional skewness up to good precision. The inset shows the trace distance norm between the PBD and numerically computed histograms as a function of $\tau_Q$. The comparison between the numerical data and the normal distribution reveals higher values of the trace norm distance.}   
    \label{fig6}
\end{figure}

\begin{figure}
    \centering
    \includegraphics[width=1\linewidth]{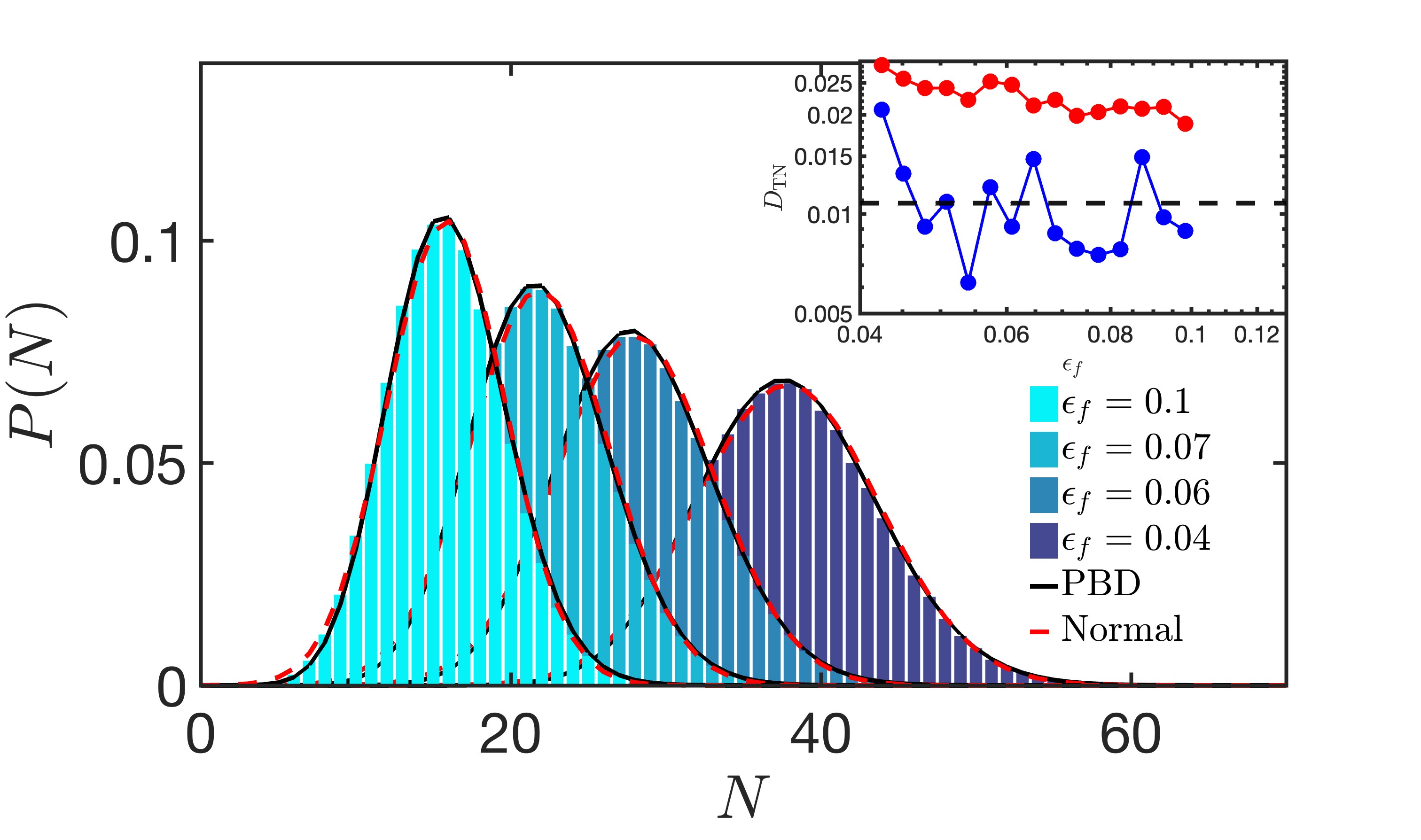}
    \caption{Histogram for the vortex-number probability distribution for fast quenches for different quench depths. Histograms are fitted to the normal approximation and the PDB. 
    The inset shows the $\epsilon_f$ dependence of the corresponding trace distance norm, indicating the accuracy of the PBD over the normal distribution.}   
    \label{fig7}
\end{figure}

Fig. \ref{fig5NEpsilon} shows the low-order cumulants of the vortex number as a function of the quench depth $\epsilon_f$ in the fats-quench limit.
The first three cumulants $\kappa_1$, $\kappa_2$ and $\kappa_3$ scale us $\kappa_1=(416\pm 10)\epsilon_f^{1.03\pm 0.01}$, $\kappa_2=(378\pm 9)\epsilon_f^{1.034\pm 0.001}$ and $\kappa_3=(294\pm 9)\epsilon_f^{1.08\pm 0.08}$.
These power-law numerical fits are in good agreement with
the beyond-KZM prediction in Eq. \eqref{cumulant_Slow} and \eqref{cumulant_Fast}, with mean-field critical exponents $\nu=1/2$ and $z=2$.
The binomial distribution approaches a normal
distribution in the large $\mathcal{N}_d$ limit
\begin{equation}\label{normal}
 P(N)=\frac{1}{\sqrt{2\pi(1-p)\langle N\rangle}} \exp \left[-\frac{(N-\langle N\rangle)^2}{2(1-p)\langle N\rangle}\right].
\end{equation}
Figure \ref{fig6} shows the histograms for the vortex-number probability distribution in slow quench region at different $\tau_Q$. Similarly, Fig. \ref{fig7} shows the histograms for the vortex-number probability distribution in the fast quench region with different $\epsilon_f$. For both slow and fast quenches, all histograms fit well to the normal distribution in Eq. \eqref{normal}, showing its robustness and lending support to the beyond-KZM prediction. However, the calculation of the trace norm distance between the numerical data and the distribution indicates the limits of this approximation, as we discuss next.

\section{Disscussion}\label{SecDiss}

To date, studies of the defect statistics beyond KZM  have focused on defect formation in one spatial dimension \cite{GomezRuiz2020,Mayo2021,Subires2022,GomezRuiz2022}, see as well \cite{delCampo2018,Cui2020,Bando2020,Bando2020R,King2022,Balducci23} in the quantum case. Studies in two spatial dimensions are scarce and have focused on other aspects,  addressing only the normal features of the defect number distribution \cite{Thudiyangal24}. Only one study to date has characterized non-normal features in a two-dimensional system consisting of a holographic superconductor with periodic boundary conditions \cite{GomezRuiz2022}. In such a setting, the vortex number distribution is confined to even outcomes; as such, it deviates from the binomial distribution and is best described as an even-Poissonian model (a limiting case of the binomial distribution conditioned to even outcomes).
Thus, evidence of the universality of the defect number distribution in two (and higher) spatial dimensions is lacking.

The number of defects in the superfluid disk with open boundary conditions is unrestricted and involves both even and odd outcomes. The non-zero values of the third cumulant and its universal scaling can be accommodated by the binomial distribution; see Eqs. (\ref{BinomDist}) and (\ref{cumulant_Slow}).
Cumulant ratios depend solely on the defect-formation probability $p$ in the binomial distribution. For the first three cumulants, these ratios are fixed as  $\kappa_2/\kappa_1=1-p$, $\kappa_3/\kappa_1=(1-p)(1-2p)$. However, the fitted power-laws are inconsistent with the binomial model, as they require values of $p$ outside the physical range $[0,1]$. 

The best-fitted distribution is instead a generalization of the binomial distributions to independent, but not identically distributed, Bernoulli trials. This is known as the Poisson binomial distribution (PBD) \cite{LeCam60,Wang93} and was predicted in Ref. \cite{GomezRuiz2020} for the most general scenario of spontaneously formed point-like defects.  Incidentally, the same distribution describes defect number distribution in quantum systems \cite{delCampo2018}, as experimentally demonstrated under unitary and open quantum dynamics \cite{Cui2020,Bando2020,King2022}.

The PBD generalizes the binomial model. To derive it in the context of beyond-KZM physics, one can still invoke the notion of a random tesselation of the superfluids by assuming that the order parameter grows independently at different locations, forming protodomains. The spatial extension of such protodomains is determined by the KZM correlation length $\hat{\xi}$. We gloss over this pictorial description being handwaving, as such protodomains are ill-defined in dimensions larger than one. 

Topological defects, such as vortices, are then expected to be formed where adjacent domains meet with a given probability, in the spirit of the geodesic rule \cite{Kibble1976}. However, the number of domains that meet at a point varies, e.g., in a random tesselation.  Two effects come into play in such a picture. First,  the frequency of occurrence of a boundary merging different domains varies with the number of them. Second, numerical simulations and experiments done by merging independent Bose-Einstein condensates show that the probability of vortex formation depends on the number of defects that meet \cite{Scherer07}.

The total number of events for defect formation $\mathcal{N}_d$ can be thus split as $\mathcal{N}_d=\sum_m\mathcal{N}_m$, where $\mathcal{N}_m=\mathcal{N}_dr_m$ accounts for the number of events for defect formation as a result of merging $m$-protodomains. Here, the probability of finding $m$ protodomains merging at a point is $r_m\geq 1$, and thus  $\sum_mr_m=1$. We further consider the conditional probability $p_m$ for a defect to be formed when $m$-protodomains meet.

The explicit form of the PBD 
\cite{Wang93}
\beqa
P(N)=\sum_{A\in F_n} \prod_{m\in A} p_m\prod_{n\in A^c}(1-p_n),
\label{mixBin}
\eeqa
involves the set $F_n$ of all subsets of $n$ integers that can be selected from ${1,\dots,\mathcal{N}}_d$, and $A^c$, the complement of $A$.
This distribution still predicts the universal scaling of all cumulants with the quench time. 
The corresponding cumulant generating function is simply the sum of the corresponding cumulant generating functions of the individual and independent Bernoulli random variables  coming as multiples of $\mathcal N_m$ for the given $m=1,2,\dots,M$:
\begin{eqnarray}
    \log \hat P(\varphi)&=&\sum_{n=1}^\infty\,\kappa_n\frac{(i\varphi)^n}{n!}\nonumber\\
    &=&\sum_{m=1}^M\,\mathcal N_m\log(1-p_m+p_m\,e^{i\varphi})
\end{eqnarray}
Thus, cumulants of $P(N)$ obey
\begin{equation}
    \kappa_n=\sum_{m=1}^M\mathcal N_m\,\kappa^{(m)}_n,
\end{equation}
where the type-$m$ cumulant is simply the cumulant of the corresponding Bernoulli variable with defect-formation probability  $p_m$ and is given by  $\kappa^{(m)}_n=(-i\partial_\varphi)^n\log\hat P(\varphi)\vert_{\varphi=0}$.  Specifically, the first few cumulants of the Poisson binomial distribution read
\begin{align}
    &\kappa_1=\mathcal N_d\sum_{m=1}^Mr_mp_m,\\
    &\kappa_2=\mathcal N_d\sum_{m=1}^Mr_mp_m(1-p_m),\\
    &\kappa_3=\mathcal N_d\sum_{m=1}^M\,r_mp_m(1-p_m)(1-2p_m),
\end{align}
with the $n$-th cumulant being
\begin{eqnarray}
    \kappa_n&=&\mathcal N_d\sum_{m=1}^M\,r_mp_m(1-p_m)\partial_{p_m}\kappa^{(m)}_{n-1}\\
    &\propto&\left(\frac{\tau_0}{\tau_Q}\right)^\frac{2 \nu}{1+z \nu}.
\end{eqnarray}    
The resulting model (\ref{mixBin}) loses predictive power with respect to the simpler binomial model due to the 
appearance of multiple parameters, i.e., the set $\{r_m,p_m\}$. However, two key features are preserved: all cumulants exhibit a universal power-law scaling as a function of the quench time, and cumulant ratios are set by constant values, given in terms of $\{r_m,p_m\}$. We further note that on idealized models of random tesselations, it is possible to go beyond and make predictions about the values of the probability $r_m$ for the merging of $m$-protodomains and the probability $p_m$ for a defect to be formed at such location using only geometric probability arguments.  

Here, we focus on the best-fitting distribution that can be determined numerically. 
The first three cumulants have three free parameters as the $r_m$ values are assumed to be independent of both $\tau_Q$ and $\epsilon_f$. Thus, the $m=2$ case with $p_1,\,p_2,\,r$ provides the simplest model beyond the binomial case to unambiguously characterize the vortex statistics, assuming two kinds of Bernoulli variables. The average number of vortices does not determine the total number of possible vortex formation points. Therefore, $r$ is used to parameterize the ratio between the events of ''type"-$1$ for vortex formation and the total average number of vortices, $p_1\mathcal N_1=r\langle N\rangle$.  Correspondingly, $p_2\mathcal N_2=(1-r)\langle N\rangle$. To find the best fitting values, the cost function $\mathcal C$ of the average square distance of the three cumulants for all given $\tau_Q$ and $\epsilon_f$ values is minimized,
\begin{widetext}
\begin{equation}
    \mathcal C=\frac{\sum_{i}\left[\left(\kappa^{\text{(num.)}}_{1,i}-\kappa^{\text{(PBD)}}_{1,i}\right)^2+\left(\kappa^{\text{(num.)}}_{2,i}-\kappa^{\text{(PBD)}}_{2,i}\right)^2+\left(\kappa^{\text{(num.)}}_{3,i}-\kappa^{\text{(PBD)}}_{3,i}\right)^2\right]}{\sum_{i}\left[\left(\kappa^{\text{(num.)}}_{1,i}\right)^2+\left(\kappa^{\text{(num.)}}_{2,i}\right)^2+\left(\kappa^{\text{(num.)}}_{3,i}\right)^2\right]}\,,
\end{equation}
\end{widetext}
where $i$ runs over all the $\tau_Q$ and $\epsilon_f$ values.
This yields the estimated parameters  $p_1=0.13,\,p_2=0.1,\,r=0.38$ with the relative mean error of the cumulants $\approx0.0011$ averaged over the $\tau_Q$ and $\epsilon_f$ values in the power-law and saturation regimes, simultaneously.  In doing so, the fitted model posits that a universal distribution function -the PBD- accommodates the universal KZM scaling, its breakdown as fast quenches, and the additional universal scaling laws as a function of the depth of the quench, i.e., the final value of the temperature.

In the plateau regime, in the range of values for $\epsilon_f$ explored, as well as in the power-law regime, as a function of $\tau_Q$, the trace norm distance between the data and the closest PBD does not exceed 0.02.

For the sake of illustration, we also performed the fitting separately for the dataset in the saturation and power law regimes and compared the corresponding errors,  summarized in the following table:
\begin{center}
\begin{tabular}{ c c c c}
 Optimization Dataset: & $\tau_Q$ & $\epsilon_f$ &  
 $\tau_Q\, {\rm and}\, \epsilon_f$\\
 $D_\mathrm{TN,\tau_Q}$&$0.0113$ & $0.0175$ & $0.0111$ \\ 
 $D_\mathrm{TN,\epsilon_f}$&$0.0111$ & $0.0158$ & $0.0108$ \\ 
 $\Delta\kappa_q(\tau_Q)$&$0.0009$&$0.0025$&$0.0008$\\
 $\Delta\kappa_q(\epsilon_f)$&$0.0034$ & $0.0007$ & $0.0023$    
\end{tabular}
\end{center}
Errors are approximately below the same value $\sim0.02$ with slightly smaller values for the cumulants for the datasets according to which the minimization itself is performed.  
Additionally, the best $\mathcal B(p,\mathcal N)$ binomial fit with $p\approx 0.08$ and $\mathcal N=\lceil \langle N\rangle/p\rceil$ based on the minimization of the trace norm results in $\sim 2.5$ times larger errors than the PBD fittings. This error analysis supports the identification of the PBD as the universal defect number distribution across all the parameter space up to numerical precision.

\section{Summary and outlook}\label{SecSumm}

The KZM is an important paradigm in nonequilibrium statistical mechanics that utilizes equilibrium properties to predict the dynamics across a quantum phase transition. It predicts the spontaneous formation of topological defects with a density that scales universally as a function of the quench time. 
Recent progress has shown that additional universal features exist in the dynamics across a phase transition that can be described in terms of equilibrium properties, such as the statistics of topological defects. However, the validity of such predictions is limited to slow quenches where the density of defects is low. Faster quenches are no longer described by KZM and show a saturation of the density of defects with the quench time. 

The recent observation that the mean defect density is also universal for fast quenches constitutes the starting point of our study.  In this work, we have established that defect statistics is universal for arbitrary values of the quench time and quench depth. Specifically, we have shown that the defect number distribution follows a Poisson binomial distribution. This is the case both at slow quenches, for which KZM holds, and for fast quenches when KZM breaks down. A testable prediction of our analysis involves the universal scaling laws of the cumulants of the defect statistics as a function of the quench time and the quench depth. 
Such tests could be implemented in experiments exploring superfluid formation in which the defect number statistics can be measured. Ultracold gases offer a relevant example, where the crossing of the superfluid phase transition can be induced by quenches of temperature, chemical potential (via a dimple trap), or interatomic interactions. In such systems, direct imaging of the vortex patterns is possible, and the vortex number distribution can be analyzed \cite{Weiler2008,Ko2019,Goo2021,Goo2022,Kim2023,lee2023observation}.
Similar studies may be at reach in other systems, such as polaritonic condensates and quantum fluids of light \cite{Carusotto13}, colloidal monolayers \cite{Keim15}, and multiferroics \cite{Lin14,Du2023}.

\acknowledgements
It is a pleasure to acknowledge discussions with 
Matteo Massaro, Seong-Ho Shinn, and Kasturi Ranjan Swain.
This work is supported by the National Natural Science Foundation of China (under Grants No. 11275233), the Postgraduate Research \& Practice Innovation Program Jiangsu Province (KYCX22\_3450), and the Luxembourg National Research
Fund (FNR), grant references 17132054 and 17132060. For the purpose of open access, the author has applied a Creative
Commons Attribution 4.0 International (CC BY 4.0) license to any Author Accepted Manuscript version arising 
from this submission.

\bibliography{ref}
\end{document}